\documentclass[preprint,3p,12pt]{elsarticle}

\usepackage{amssymb}
\usepackage{amsmath}
\usepackage{mathtools}
\usepackage{amsfonts}
\usepackage{amsbsy}
\usepackage{color}
\usepackage{float}
\usepackage{graphicx}
\usepackage{caption}
\usepackage{subcaption}
\usepackage{lineno}
\usepackage{comment}
\usepackage{algorithm}
\usepackage{algorithmic}
\usepackage{tcolorbox}
	
\newcommand{\bs}{\boldsymbol}

\newcommand{\bm}{\mathbf}
\newcommand{\mrm}{\mathrm}
\newcommand{\domain}[1][]{
	\ensuremath{\mathcal{H}^{#1}}}

\newcommand{\state}{\underline}
\newcommand{\base}[1][\boldsymbol{\xi}]{\langle#1\rangle}

\begin{document}

\begin{frontmatter}

\title{A Message Passing Neural Network Surrogate Model for Bond-Associated Peridynamic Material Correspondence Formulation}

\author[ucb]{Xuan Hu}

\author[ucb]{Qijun Chen}

\author[rvh]{Nicholas H. Luo}

\author[mhs]{Richy J. Zheng}

\author[ucb]{Shaofan Li\corref{cor1}}
\ead{shaofan@berkeley.edu}
\cortext[cor1]{Corresponding author}

\address[ucb]{Department of Civil and Environmental Engineering, University of California, Berkeley, CA 94720, USA}
\address[rvh]{San Ramon Valley High School, CA 94526, USA}
\address[mhs]{Miramonte High School, CA 94563, USA}

\begin{abstract}
Peridynamics is a non-local continuum mechanics theory that offers unique advantages for modeling problems involving discontinuities and complex deformations. Within the peridynamic framework, various formulations exist, among which the material correspondence formulation stands out for its ability to directly incorporate traditional continuum material models, making it highly applicable to a range of engineering challenges. A notable advancement in this area is the bond-associated correspondence model, which not only resolves issues of material instability but also achieves computational accuracy comparable to finite element analysis (FEA). However, the bond-associated model typically requires higher computational costs than FEA, which can limit its practical application.\\
To address this computational challenge, we propose a novel surrogate model based on a message-passing neural network (MPNN) specifically designed for the bond-associated peridynamic material correspondence formulation. Leveraging the similarities between graph structure in computer science theory and the neighborhood connectivity inherent to peridynamics, we construct an MPNN that can transfers domain knowledge from peridynamics into a computational graph and shorten the computation time via GPU acceleration. Unlike conventional graph neural networks that focus on node features, our model emphasizes edge-based features, capturing the essential material point interactions in the formulation. Additionally, an attention mechanism is integrated into the MPNN to enhance its representation of bond-associated dynamics, ensuring that significant interactions are weighted appropriately. A key advantage of this neural network approach is its flexibility: it does not require fixed neighborhood connectivity, making it adaptable across diverse configurations and scalable for complex systems. Furthermore, the model inherently possesses translational and rotational invariance, enabling it to maintain physical objectivity—a critical requirement for accurate mechanical modeling.\\
We validate the accuracy and efficacy of this surrogate model through several numerical examples, demonstrating its potential as a powerful tool for efficient and accurate peridynamic simulations. This work opens new avenues for applying peridynamic models in computationally demanding scenarios, providing a viable alternative to traditional methods with the potential for significant computational savings.
\end{abstract}


\end{frontmatter}
\section{Introduction}
Peridynamics is a nonlocal continuum mechanics theory that addresses the limitations of the classical local theory in dealing with spatial discontinuities and accounting for length scale effects~\cite{silling2000bond, silling2007state,bobaru2012horizon,bobaru2016handbook,chen2020higher,chan2023higher}. The development of peridynamics began with the seminal work by Silling~\cite{silling2000bond} on reformulation of elasticity theory for discontinuities and long-range forces, where pairwise bond-based interactions within finite distance called horizon are formulated. In this bond-based formulation, the force density of a bond depends only on its stretch. While it is effective in capturing fracture phenomena, the bond-based formulation is limited in describing general material behaviors such as arbitrary Poisson ratio and nonlinear constitutive relationship, due to the usage of a pairwise potential that is totally independent of all other local conditions~\cite{silling2007state}. To overcome this limitation, the state-based formulation which leveraged the concept of state to rewrite the material-dependent part of the peridynamic model was introduced~\cite{silling2007state}. More importantly, the material correspondence formulation, a subset of the state-based formulations, bridges the gap between peridynamics and the classical continuum mechanics theory by allowing direct incorporation of continuum material models into peridynamics. This is achieved by introducing nonlocal deformation gradient and stress tensors in a manner equivalent to the classical continuum mechanics theory but within a nonlocal framework. However, the material correspondence formulation is not without challenges. One well-known issue of the formulation is the existence of material instability or zero-energy modes manifested in the form of oscillation in the displacement field. These modes arise when certain deformation states do not contribute to the strain energy, leading to non-physical solutions and numerical instabilities. Among existing strategies proposed in the literature to address this issue, the bond-associated formulations are the most effective and provide more accurate accounting of bond-level quantities such as deformation gradient and stresses. Chen et al.~\cite{chen2023influence, chen2024generalized} proposed a family of non-spherical influence function and developed the corresponding material correspondence model to improve the accuracy of bond-level quantities such as deformation gradient and stress. Unlike the conventional formulation, where the influence function is spherical and depends only on the bond length, the proposed non-spherical influence functions take into account both the bond length and the bond relative angle (with respect to a target bond). This novel bond-associated correspondence model has achieved great success in inherently eliminating the material instability in the conventional formulation. However, the computational cost of this bond-associated formulation is higher than conventional ones, which is caused by iterative calculation of the non-spherical influence function. One way to overcome this issue is using GPU acceleration technique, which is commonly adopted in artificial intelligence (AI) or machine learning (ML) models.\\
Graph neural networks (GNNs) are a type of neural network architecture that can operate on graph structures, such as those found in social networks, molecules, and materials. Unlike traditional neural networks, which are designed for tabular data, GNN models can be proposed to operate directly on the  graphs with arbitrary edges, giving them the flexibility to train task-specific representations more relevant to the properties of interest. More specifically, GNN captures the dependence of graphs via message-passing between the nodes or edges of graphs. According to the pattern of message-passing and aggregation, it contains different variants, such as graph recurrent network (GRN) \cite{ruiz2020gated}, graph convolutional network (GCN) \cite{kipf2016semi}, graph attention network (GAT) \cite{velivckovic2017graph}, etc. As a sub-class of GNN, message-passing neural networks (MPNNs) have recently shown great success in a wide range of applications\cite{tang2023application}. Due to the similarities existed between non-locality of peridynamics and graph structure, bond-associated influence functions and attention mechanism, it is possible to build a MPNN-based surrogate model for the bond-associated material correspondence model. By doing so, we can not only achieve faster running speed by leveraging GPU computation, but also pave the path of discovering new material response through solving inverse problems. Some attempts have been made by incorporating peridynamics with the novel AI/ML methods. Ning et al.\cite{ning2023peridynamic}, Madenci et al.\cite{madenci2022peridynamics, haghighat2021nonlocal} developed physics-informed neural network based on peridynamic differential operator. Yu et al.\cite{yu2024nonlocal} proposed energy-informed neural network as a surrogate model for computing the displacement, which essentially is also a physics-informed neural network. Jafarzadeh et al.\cite{jafarzadeh2024peridynamic} introduced the peridynamic neural operator framework which learns ordinary state-based peridynamic constitutive models from data. So far, few attempts for combining GNNs with peridynamic material correspondence formulation have been reported in the literature. Due to their similarities in both structures and mechanisms, we believe this can be a promising direction for AI-assisted scientific computing.
The remaining part of this paper is organized as follows: a brief introduction to bond-associated peridynamics correspondence formulation is presented in section 2; the proposed message passing neural network framework is given in section 3.
\section{Bond-Associated Peridynamics Correspondence Formulation}
\subsection{Definition and Notations}
\begin{figure}
    \centering
    \includegraphics[width=0.7\linewidth]{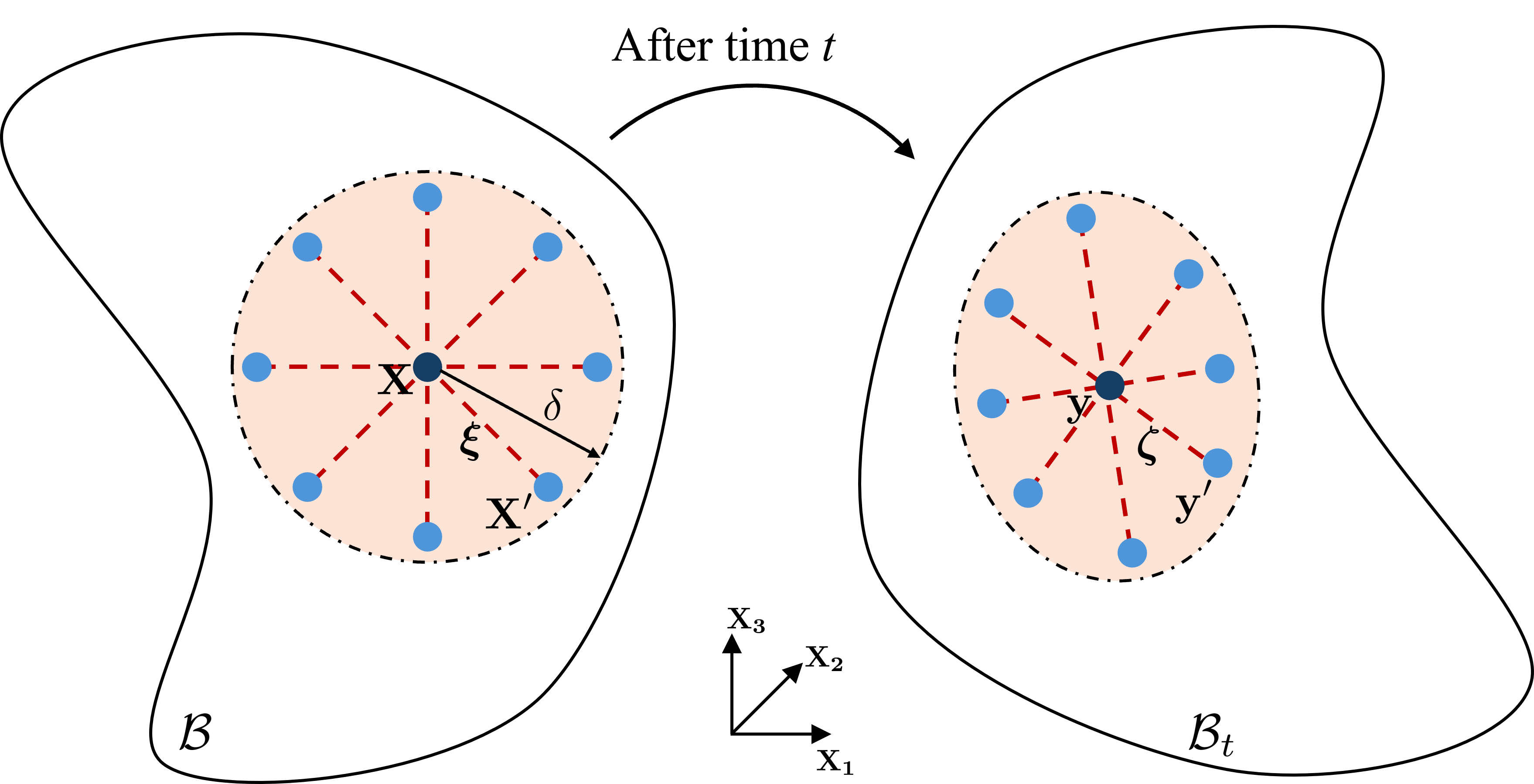}
    \caption{Schematic diagram of peridynamics model}
    \label{fig:pd_diagram}
\end{figure}

 As shown in Fig. \ref{fig:pd_diagram}, the geometric domain of interest in referential configuration $\mathcal{B}$ is modeled as an assembly of material points with volume. For each material point $\bm X$, it interacts with its neighboring material points located within a Euclidean distance $\delta$, which is known as \textit{horizon}, through nonlocal interactions as bond forces. The points within the horizon is called \textit{neighbors} and the collection of all neighbors is referred to as \textit{neighborhood}, denoted as $\domain$. The \textit{relative position in reference configuration} $\mathcal{B}$ between a material point $\bm X$ and its neighbor $\bm X'$ is referred to as a bond $\bs\xi = \bm X' - \bm X$. Let $\bm y(\bm X, t)$ represents the corresponding position of the material point $\bm X$ in the current configuration $\mathcal{B}_t$, after time $t\geq 0$. The \textit{relative position in current configuration} $\mathcal{B}_t$ between two neighboring points is denoted as $\bs\zeta = \bm y'(\bm X', t) - \bm y(\bm X, t)$.\\
 The development of the peridynamic material correspondence formulation introduces the concept of state~\cite{silling2007state}. A \textit{state} of order $m$ is a function $\state{\bm A}\base[\bullet]:\domain\rightarrow \mathcal{L}_m$, which maps a vector in the neighborhood $\domain$ to the tensor space  $\mathcal{L}_m$ of order $m$. For instance, if $m=0$, a bond is mapped to a scalar space $\mathcal{L}_0$, and this is referred to as a scalar state. Scalar states are usually written in lowercase, non-boldface with an underscore, such as $\state{\omega}\base, \state{\mrm{w}}\base$. If $m=1$, a bond is mapped to a vector space $\mathcal{L}_1$, and the state is called vector state. vector states and other states of order $m\geq 1$ are conventionally written in uppercase boldface with an underscore, such as $\state{\bm y}\base$ or $\state{\bm A}\base$. According to this definition, a state can be readily identified that maps the initial bond vector $\bs\xi$ to the current bond vector $\bs\zeta$. This is termed the \textit{deformed state} and denoted by $\state{\bm y}\base$. For a bond connecting a material point $\bm X$ and one of its neighbors at any time $t$, the deformed state is $\state{\bm y}[\bm X, t]\base = \bm y'(\bm X', t) - \bm y(\bm X, t) = \bs\zeta$. For clarity, square brackets are introduced to indicate the spatial or temporal information on which a state depends, e.g., $[\bm X, t]$, while parentheses are adopted to denote all other quantities that a state depends on, e.g., $(\bs\xi, \bs\xi')$. In addition, the standard convention in classical continuum mechanics theory is followed in this paper. Variables with uppercase subscripts, such as $X_I$, refer to the components defined in reference configuration $\mathcal{B}$ while the ones with lowercase subscripts, such as $y_j$ denote the components in current configuration $\mathcal{B}_t$. Einstein summation notation is also employed here to facilitate the representation of complex tensor operations. Boldface letters indicate vectors or tensors.
\subsection{Nonlocal Deformation Gradient}
In authors' previous research, the deformation gradient associated with bond $\bs\xi^{IJ}$ writes as
\begin{align}
    \bm F^{IJ} = \left[\sum_{L\in\domain[I]} \state\omega(\bs\xi^{IJ},\bs\xi^{IL})\state{\bm y} ^{IL}\otimes \bs\xi^{IL}V^L\right]\left(\bm K^{IJ}\right)^{-1}
    \label{eq:def_F}
\end{align}
with 
\begin{align}
    \bm K^{IJ} = \sum_{L\in\domain[I]} \state\omega(\bs\xi^{IJ}, \bs\xi^{IL}) \bs\xi^{IL} \otimes \bs\xi^{IL}V^L,
    \label{eq:def_K}
\end{align}
where $\domain[I]$ refers to the neighborhood of particle $I$; $J, L$ are both the neighbors within $\domain[I]$; $\bs\xi^{IJ}:=\bm X^J - \bm X^I, \bs\xi^{IL}:=\bm X^L - \bm X^I$ are the bonds connecting material points $I$ and $J$, $I$ and $L$ respectively; $\state{\bm y}^{IL}$ is the deformed state of $\bs\xi^{IL}$ in the current configuration $\mathcal{B}_t$; $V^L$ represents the discretized volume that associated with the particle $L$; and $\state{\omega}(\bs\xi^{IJ}, \bs\xi^{IL})$ is the influence function that is associated to bond $\bs\xi^{IJ}$, which satisfies
\begin{align}
    &\state\omega(\bs\xi^{IJ}, \bs\xi^{IL}) \propto \exp\left(-n_1\frac{||\bs\xi^{IJ}| - |\bs\xi^{IL}||}{\delta}\right)\left(\frac{1}{2} + \frac{1}{2}\cos(\widehat{\bs\xi^{IJ}\bs\xi^{IL}})\right)^{n_2},\nonumber\\
    &\mrm{and}\quad\sum_{L\in\domain[I]} \state\omega(\bs\xi^{IJ}, \bs\xi^{IL}) V^L = 1.
    \label{eq:non-spherical}
\end{align}
In Eq. \eqref{eq:non-spherical}, $\exp(\cdot)$ is the exponential function, $n_1, n_2$ are controlling parameters that can control the shape of the influence function over the neighborhood $\mathcal{H}^I$; $\widehat{\cdot}$ indicates angle between two bonds.
\subsection{Force Density State}
The force density state corresponding to the unified bond-associated deformation gradient can be derived following the same procedure as outlined by Silling et al.~\cite{silling2007state}. Let's define the nonlocal strain energy density for bond-associated models as:
\begin{equation}
    \mathcal{W} = \sum_{J\in\domain[I]}\state{\mrm{w}}\base[\bs\xi^{IJ}] \bm P^{IJ}:\bm F^{IJ} V^J,
\end{equation}
where $\state{\mrm{w}}\base[\bs\xi^{IJ}]$ is a scalar state-valued weight function that satisfies $\sum_{J\in\domain[I]}\state{\mrm{w}}\base[\bs\xi^{IJ}]V^J=1$; $\bm F^{IJ}$ denotes the deformation gradient associated with bond $\bs\xi^{IJ}$, and $\bm P^{IJ}$ is the first Piola–Kirchhoff stress (PK1 stress) corresponding to $\bm F^{IJ}$.\\
Assuming $\bm F^{IJ}$ is differentiable, the Fr\'echet derivative of $\bm F^{IJ}$ can be obtained using the following equation:
\begin{align}
    \bm F^{IJ}(\state{\bm y}+\Delta\state{\bm y}) =& \left[\sum_{L\in\domain[I]}\state\omega\cdot(\state{\bm y}^{IL} + \Delta\state{\bm y}^{IL})\otimes \bs\xi^{IL} V^L\right]\cdot\left(\bm K^{IJ}\right)^{-1}\nonumber\\
    =& \left[\sum_{L\in\domain[I]}\state\omega\cdot\state{\bm y}^{IL}\otimes \bs\xi^{IL} V^L\right]\cdot\left(\bm K^{IJ}\right)^{-1} + \nonumber\\
    &\left[\sum_{L\in\domain[I]}\state\omega\cdot\Delta\state{\bm y}^{IL}\otimes \bs\xi^{IL} V^L\right]\cdot\left(\bm K^{IJ}\right)^{-1}
    \label{eq:F_y_dy}
\end{align}
From the Fr\'echet derivative provided in Eq.~\eqref{eq:F_y_dy} and by letting $\bm F^{IJ}(\state{\bm y} + \Delta\state{\bm y}) = \bm F^{IJ} + \Delta\bm F^{IJ}$, the increment of deformation gradient $\Delta \bm F^{IJ}$ can be expressed as
\begin{align}
    \Delta \bm F^{IJ} = \left[\sum_{L\in\domain[I]}\state\omega\cdot\Delta\state{\bm y}^{IL}\otimes \bs\xi^{IL} V^L\right]\cdot\left(\bm K^{IJ}\right)^{-1}
\end{align}
Similarly, the increment of the nonlocal strain energy density $\Delta\mathcal{W}$ due to $\Delta\state{\bm y}$ can be derived using Fr\'echet derivative as:
\begin{align}
    \Delta\mathcal{W} = & \sum_{J\in\domain[I]} \state{\mrm{w}}\base[\bs\xi^{IJ}]\bm P^{IJ}:\Delta\bm F^{IJ}V^J\nonumber\\
    =& \sum_{J\in\domain[I]} \state{\mrm{w}}\base[\bs\xi^{IJ}]\bm P^{IJ}:\left[\left(\sum_{L\in\domain[I]}\state\omega\cdot\Delta\state{\bm y}^{IL}\otimes \bs\xi^{IL} V^L\right)\cdot\left(\bm K^{IJ}\right)^{-1}\right]V^J\nonumber\\
    =& \sum_{L\in\domain[I]} \left(\sum_{J\in\domain[I]}\state{\mrm{w}}\base[\bs\xi^{IJ}]\state\omega \bm P^{IJ}\left(\bm K^{IJ}\right)^{-1}V^J\right)\bs\xi^{IL}\cdot \Delta\state{\bm y}^{IL} V^L\nonumber\\
    =& \sum_{J\in\domain[I]} \left(\sum_{L\in\domain[I]}\state{\mrm{w}}\base[\bs\xi^{IL}]\state\omega \bm P^{IL}\left(\bm K^{IL}\right)^{-1}V^L\right)\bs\xi^{IJ}\cdot \Delta\state{\bm y}^{IJ} V^J
\end{align}
According to the work conjugate relation, the corresponding force density state in the bond-associated correspondence formulation is obtained as
\begin{align}
    \state{\bm T}^{IJ} = \state{\bm T}\base[\bs\xi^{IJ}] = \left( \sum_{L\in\domain[I]}\state\omega(\bs\xi^{IJ}, \bs\xi^{IL}) \state{\mrm{w}}\base[\bs\xi^{IL}]\bm P^{IL}\left(\bm K^{IL}\right)^{-1}V^L\right)\bs\xi^{IJ}
    \label{eq:force_state}
\end{align}
\subsection{Governing Equation}
The equation of motion for the bond-associated material correspondence models is the same as that for the conventional model proposed by Silling et al.~\cite{silling2007state}. The equation of motion is expressed as
\begin{align}
    \rho(\bm X^I)\Ddot{\bm u}(\bm X^I, t) = \sum_{J\in\domain[I]}\left(\state{\bm T}^{IJ} - \state{\bm T}^{JI}\right)V^J + \bm b(\bm X^I, t),
    \label{eq:eom}
\end{align}
where $\rho(\bm X^I)$ is the mass density of material point $\bm X^I$; $\Ddot{\bm u}(\bm X^I, t) := \partial^2\bm u / \partial t^2$ represents the second order derivative of displacements $\bm u^I$ with respect to time $t$; and $\bm b(\bm X^I, t)$ is the body force density of material point $\bm X$. The force density $\state{\bm T}^{IJ}$ can be computed by Eq. \eqref{eq:force_state}. Similarly, $\state{\bm T}^{JI}$ can be expressed as
\begin{align}
    \state{\bm T}^{JI} = \state{\bm T}\base[\bs\xi^{JI}] = \left( \sum_{L\in\domain[J]}\state\omega(\bs\xi^{JI}, \bs\xi^{JL}) \state{\mrm{w}}\base[\bs\xi^{JL}]\bm P^{JL}\left(\bm K^{JL}\right)^{-1}V^L\right)\bs\xi^{JI}
    \label{eq:force_state2}
\end{align}
Since most quantities in Eq. \eqref{eq:force_state2} are evaluated at material point $J$ instead of $I$, $\state{\bm T}^{IJ}$ and $\state{\bm T}^{JI}$ may not have the same magnitude or parallel to each other.
\section{Message Passing Neural Network}
\subsection{Invariance, Equivariance and Momentum Conservation Laws}
Considering a physical system with $\bm z\in\mathbb{R}^{d}$ as the initial position and $\bm f$ as the resulting mechanical response, this system should be invariant to translation and equivariant to rotation\cite{liu2023ino}. Translational invariance means that translating $\bm z$ by a constant vector $\bm g$ will result in the same mechanical response, i.e. 
\begin{align}
    \bm f\base[\bm z + \bm g] = \bm f\base[\bm z]
\end{align}
According to Noether's theorem\cite{noether1971invariant}, this guarantees the conservation of linear momentum. Rotational equivariance implies that the response $\bm f$ remains unchanged when the initial position $\bm z$ is subjected to rigid body rotation. In mathematics, it can be expressed as
\begin{align}
    \bm f\base[\bm R\bm z] = \bm R\bm f\base[\bm z]
\end{align}
where $\bm R$ is a orthonormal matrix. According to Noether's theorem\cite{noether1971invariant}, this guarantees the conservation of angular momentum.
\subsection{Construction of Neural Network}
\begin{figure}
    \centering
    \begin{subfigure}[t]{0.7\linewidth}
        \centering
        \includegraphics[width=\linewidth]{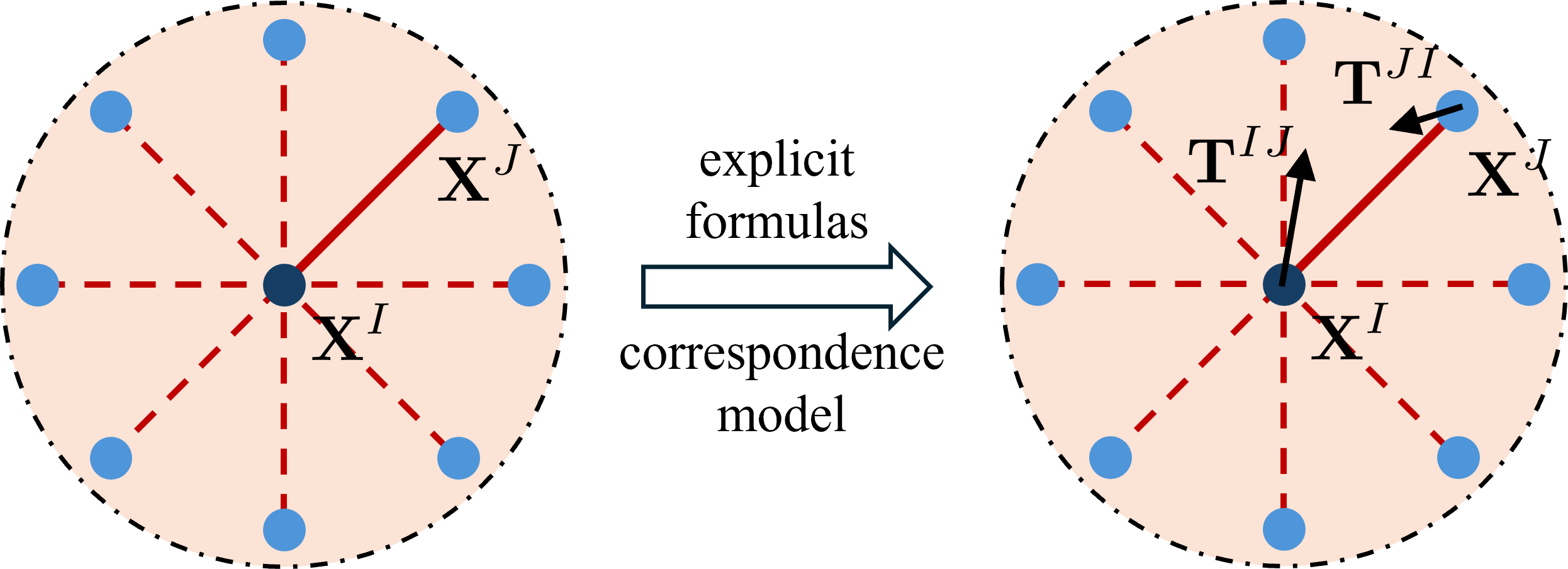}
        \begin{center}
            (a)
        \end{center}
    \end{subfigure}
    \begin{subfigure}[t]{0.7\linewidth}
        \centering
        \includegraphics[width=\linewidth]{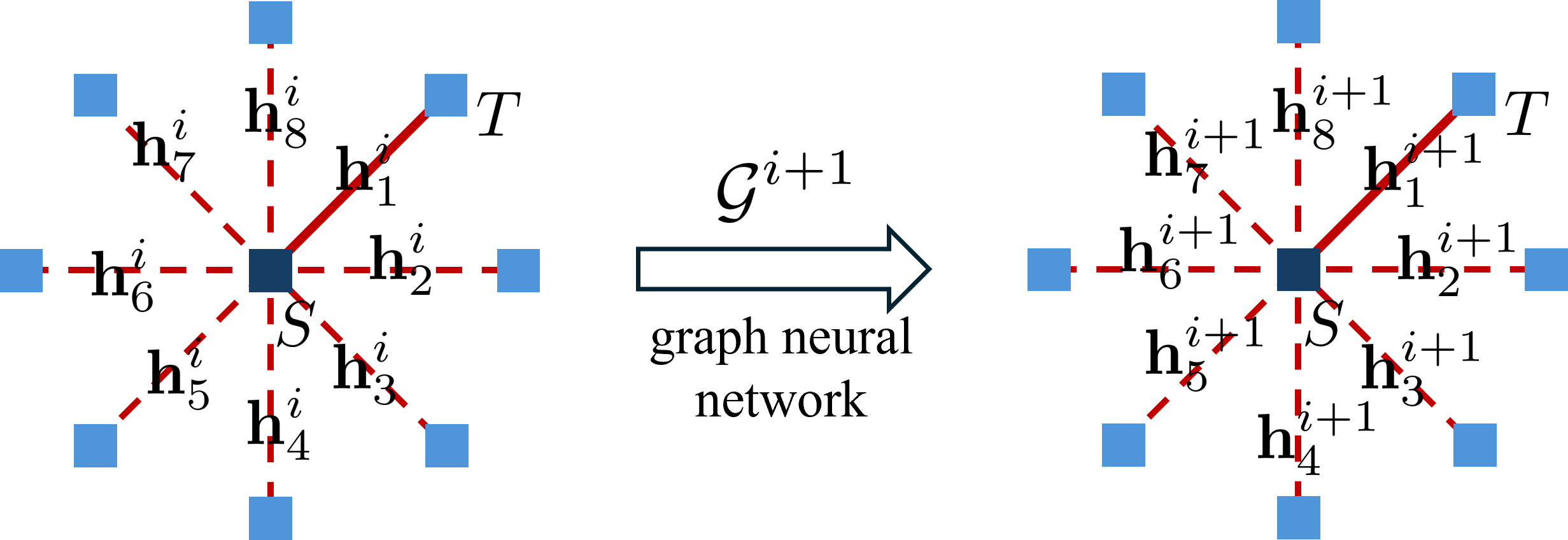}
        \begin{center}
            (b)
        \end{center}
    \end{subfigure}
    \caption{Schematic diagram of (a) peridynamic correspondence model and (b) MPNN model}
    \label{fig:pd_mpnn_diagram}
\end{figure}
As shown in Fig. \ref{fig:pd_mpnn_diagram}, the peridynamics correspondence model can be directly transformed into a graph structure, where material points $\bm X^I, \bm X^J$ become source and target nodes $S, T$ respectively while the bonds connecting $X^I$ and its neighborhood turn to edges. Let $\bm h_{ab}^i$ be the features on the edge ab no the hidden layer $i$, the following equation defines the edge-wise message-passing feature update in hidden layer $i+1$:
\begin{align}
    \bm h_{ab}^{i+1} = \mathcal{G}^{i+1}\left(\bm h_{ab}^i,\underset{c\in\domain[a]}{\bigoplus}\phi^{i+1}\left(\bm h_{ab}^i, \bm h_{ac}^i\right)\right)
\end{align}
where $\bm h_{ab}^{i+1}$ refers to the latent feature on edge ab in hidden layer $i+1$; $\oplus$ denotes a differentiable, permutation invariant function such as sum, mean or max; and $\mathcal{G}$ and $\phi$ represent differentiable functions such as multi-layer perceptrons (MLPs). After constructing multiple layers, we can eventually map the initial graph features $\bm h^0_{ab}$ to the bond-associated force states $\hat{\bm T}\base[\bs\xi^{ab}]$, i.e. 
\begin{align}
    \widehat{\bm T}\base[\bs\xi^{ab}] = \mathcal{G}^n\circ\mathcal{G}^{n-1}\circ...\circ\mathcal{G}^1 \left(\bm h_{ab}^0,\underset{c\in\domain[a]}{\bigoplus}\phi^{1}\left(\bm h_{ab}^0, \bm h_{ac}^0\right)\right)
\end{align}
To satisfy translation invariance, we select
\begin{align}
    \bm h_{ac}^0 := [\bs\xi^{ac},~~\state{\bm y}^{ac}].
\end{align}
\textit{proof of translational invariance and rotational equivariance}: Supposing all the material points $\bm y^J$ within neighborhood $\domain[I]$ are transformed as $\bm y^J_{new} = \bm R\bm y^J + \bm g$, the second component in feature $\bm h_{ac}^0$ becomes
\begin{align}
    \state{\bm y}^{ac}_{new} =& \bm y^c_{new} - \bm y^a_{new}\nonumber\\
    =& \bm R\bm y^c + \bm g - \bm R\bm y^a - \bm g\nonumber\\
    =& \bm R\state{\bm y}^{ac}
\end{align}
Thus the translational invariance and rotational equivariance is satisfied.
Since this is a surrogate model for computing the force state in peridynamic framework, the loss function is defined as
\begin{align}
    \mathcal{L} = \frac{1}{N_b}\sum_{i=1}^{N_b} \lVert\widehat{\bm T}\base[\bs\xi^{ab}] - \state{\bm T}\base[\bs\xi^{ab}]\rVert
\end{align}
where $N_b$ represents the total number of bonds within the domain $\mathcal{B}$, and $\lVert\bullet\rVert$ implies the second norm of the vector.
\newpage
\bibliographystyle{unsrt}
\bibliography{ref}
\end{document}